\documentclass[aps,prc,twocolumn,groupedaddress]{revtex4} 
 
\usepackage{graphicx,color} 
\usepackage{bm}
\usepackage{amsmath,amssymb,amsfonts}
\usepackage{amssymb}
\usepackage{ulem}

\def\bea{\begin{eqnarray}}
\def\eea{\end{eqnarray}}
\def\be{\begin{equation}}
\def\ka{\kappa}
\def\la{\lambda}
\def\ee{\end{equation}}

\begin{document}

\title{Particle-particle random phase approximation applied to Beryllium isotopes}

\author{G. Blanchon$^{(1)}$, N. Vinh Mau$^{(1,2)}$, A. Bonaccorso$^{(3)}$, M. Dupuis$^{(1)}$, and N. Pillet$^{(1)}$}
\affiliation{$^{(1)}$ CEA,DAM,DIF F-91297 Arpajon, France}
\affiliation{$^{(2)}$ Institut de Physique Nucl\'eaire, IN2P3-CNRS, Universit\'e Paris-Sud, F-91406, Orsay Cedex, France\\}
\affiliation{$^{(3)}$ Istituto Nazionale di Fisica Nucleare, Sezione di Pisa, Largo Pontecorvo 3, 56127 Pisa, Italy}

\begin{abstract}
{This work is dedicated to the study of even-even $^{8-14}$Be isotopes using the particle-particle Random Phase Approximation that 
accounts for two-body correlations in the core nucleus. A better description of energies and two-particle amplitudes is obtained in comparison with models assuming a neutron closed-shell (or subshell) core. A Wood-Saxon potential corrected by a phenomenological particle-vibration coupling term has been used
for the neutron-core interaction and the D1S Gogny force for the neutron-neutron interaction. Calculated ground state properties as well as excited state ones are discussed and compared to experimental data. In particular, results suggest the same 2s$_{1/2}$-1p$_{1/2}$ shell inversion in $^{13}$Be as in $^{11}$Be.}
\end{abstract}

\maketitle

\section{Introduction}
\label{intro}

In two previous studies \cite{labiche_99,pacheco_02}, it has been proposed the same inversion in $^{13}$Be between 2s$_{1/2}$ and 1p$_{1/2}$ shells as the one found in $^{11}$Be \cite{tanihata_85} and $^{10}$Li \cite{chartier_01}. This assumption has been suggested to obtain a good description of $^{14}$Be two-neutron separation energy. 
In Ref.\cite{labiche_99}, a simple pairing model was utilized whereas in Ref.\cite{pacheco_02} a particle-particle RPA (pp-RPA) approach with the D1S Gogny force \cite{berger_91,decharge_80} was introduced.

The present study is an extension of the work of Pacheco and Vinh Mau \cite{pacheco_02}. It aims at getting additional and essential information on the structure of $^{13}$Be from the study of $^{12}$Be and $^{14}$Be. Recently, several  measurements on Beryllium isotopes and especially on their excited states have been performed \cite{shimoura_03,lecouey_04,simon_04,simon_07,sugimoto_07,kondo,kanungo_10}. These new experimental data associated with our present calculations are important both to assess the validity of the model itself for the description of excited states and to test with more constraints the hypothesis of the inversion between 2s$_{1/2}$ and 1p$_{1/2}$ shells in $^{13}$Be. The pp-RPA approach is a three-body model that provides information on $(A-2)$, $A$ and $(A+2)$ nuclei starting from a reference core nucleus with $A$ nucleons. A spherical symmetry is assumed. In the present work, we apply it both to $^{10}$Be and $^{12}$Be cores in order to get consistent information on $^{8-14}$Be isotopes. We use a Wood-Saxon (WS) potential corrected by a phenomenological particle-vibration coupling \cite{vinhmau_95,labiche_99,pacheco_02} for the neutron-core interaction and the D1S Gogny force to describe the neutron-neutron effective interaction. Moreover, in the previous study of Pacheco and Vinh Mau \cite{pacheco_02}, the two-body spin-orbit interaction term was neglected in the study of 0$^{+}_{1}$ ground state \cite{girod_83}. As in the present work we are interested in excited states with different spins and parities, the spin-orbit term is included. 
 
In this article, results associated with ground state properties of even-even $^{8-14}$Be isotopes as well as their excited spectra and transition probabilities are 
presented. Comparisons with experimental data are also discussed in detail. In Sec.\ref{ppRPA1}, we briefly introduce the pp-RPA approach and give a few analytical formulas concerning radii and transition probabilities. Sec.\ref{result10Be} is devoted to results obtained from a $^{10}$Be core. The neutron-$^{10}$Be interaction has been fitted from experiment \cite{ajzenberg_90} and used to generate the single particle basis. The ability of the model to reproduce the experimental knowledge on $^{10}$Be and $^{12}$Be is demonstrated. In Sec.\ref{result12Be}, similar analysis with a $^{12}$Be core is presented. As the neutron-$^{12}$Be interaction is not known precisely from experiment, we investigate different scenarii for the neutron-core interaction, constrained by experimental knowledge on $^{12}$Be, $^{13}$Be and $^{14}$Be. Conclusions are given in Sec.\ref{concl}.

\section{pp-RPA formalism and associated observables}
\label{ppRPA1}

The pp-RPA is a well-known formalism used to study nuclei which can be approximated as a core plus or minus two nucleons. An important property of this approach is its ability to account for two-body correlations in the core nucleus. Different ways of deriving pp-RPA equations can be found in the literature \cite{ring}. For example,
the Green's function method shows that pp-RPA approach introduces in the core ground state contribution of multiparticle-multihole configurations \cite{mulhall_67}. 

In this part, we recall briefly the standard equations in order to introduce our notations. Starting from a nucleus with $A$ nucleons, the pp-RPA equations describing the 
nuclei with $(A+2)$ and $(A-2)$ nucleons are,
\begin{multline}
 \left(\Omega -[\epsilon_{a}+\epsilon_b]\right) x_{ab} -\sum_{kl}\langle kl|V|\widetilde{ab}\rangle x_{kl}\\
  -\sum_{\kappa \lambda}\langle \kappa \lambda|V|\widetilde{ab}\rangle x_{\kappa\lambda} = 0,\label{rpa1}
\end{multline}
\begin{multline}
 \left(\Omega -[\epsilon_{\alpha}+\epsilon_{\beta}]\right)x_{\alpha \beta}+
\sum_{kl}\langle kl|V|\widetilde{\alpha \beta}\rangle x_{kl}\\
  +\sum_{\kappa \lambda}\langle \kappa \lambda|V|\widetilde{\alpha \beta}\rangle x_{\kappa \lambda}=0. \label{rpa2}
\end{multline}
In Eqs.(\ref{rpa1}) and (\ref{rpa2}), the set of $\epsilon$ are single particle energies determined together with the corresponding wave functions in a given neutron-core potential. The chosen potential is presented in Sec.\ref{result10Be} (see Eq.(\ref{ws})). Matrix elements of the two-body interaction $V$ are antisymmetrized. Latin indices stand for unoccupied single particle orbits and greek indices for occupied ones. The quantities $x$ gives the standard two-nucleon amplitudes $X$ and $Y$ for nuclei with $(A+2)$ and $(A-2)$ nucleons, respectively. They read 
\bea
X_{mn}^{(N)}&=&\langle A+2,N|{\cal{A}}_{mn}^{\dagger}|A,\tilde 0 \rangle \label{ee1}, \\
Y_{mn}^{(M)}&=&\langle A-2, M|{\cal{A}}_{mn}|A,\tilde 0\rangle, \label{ee2}
\eea
where $m\leq n$. The states of $(A+2)$ or $(A-2)$ nuclei  are labeled by $N$ or $M$, respectively. The state $|A,\tilde 0\rangle$ stands for the correlated 
ground state of the core. $|A+2,N\rangle$ and $|A-2,M\rangle$ are the states of the $(A+2)$ and $(A-2)$ nuclei, respectively. In Eq.(\ref{ee1}), the pair creation 
operator ${\cal{A}}^{\dagger}$ is defined as  
\be
{\cal{A}}_{ab}^{\dagger}(J,M_J)=\nu_{ab} \sum_{m_a,m_b} \left\langle j_{a} j_{b} m_{a} m_{b}|J,M_J \right\rangle a_{a}^{\dagger}
a_{b}^{\dagger},  \label{aa1}
\ee
with $a \leq b$ and $\nu_{ab}=\left(1+\delta_{j_{a}j_{b}}\right)^{-{1/2}}$ with the two nucleons in the same spherical j-orbital. The annihilation 
operator ${\cal{A}}$ in Eq.(\ref{ee2}) is deduced from Eq.(\ref{aa1}). These expressions are valid for both occupied and unoccupied states. \\
In Eqs.(\ref{rpa1}) and (\ref{rpa2}), $\Omega$ is the energy of the state related to $(A+2)$ or $(A-2)$ nucleus taking as reference the $A$ nucleus 
ground state energy. Hence, for the $(A+2)$ nucleus,
\bea
E_{N}(A+2)-E_{0}(A)=\Omega_{N},\label{e}\\
X_{ab}^{(N)}=x_{ab}^{(N)}, \ \ X_{\alpha \beta}^{(N)}=x_{\alpha\beta}^{(N)},
\eea
and for the $(A-2)$ nucleus,
\bea
E_{M}(A-2)-E_{0}(A)=-\Omega_{M} , \label{e2}\\
Y_{ab}^{(M)}=x_{ab}^{(M)}, \ \ Y_{\alpha \beta}^{(M)}=x_{\alpha \beta}^{(M)}.
\eea
More details concerning the pp-RPA formalism are given in Appendix \ref{pprpa}.\\
In addition to energies and amplitudes, other observables as $rms$ radii and transition probabilities will be discussed in Sec.\ref{result10Be} and \ref{result12Be}. 
We give here the expressions for those two quantities. Details can be found in Appendices \ref{radius} and \ref{trans}. Following Ref.\cite{vinhmau_96}, the $rms$ radius $\langle \textbf{r}^{2} \rangle_{A+2}^{1/2}$ of the $(A+2)$ system can be expressed in terms of the radius of the core $\langle \textbf{r}^{2} \rangle_{A}^{1/2}$,
\be
\langle \textbf{r}^{2}\rangle_{A+2}={{A}\over{A+2}}\langle \textbf{r}^{2}\rangle_{A}+\delta\langle \textbf{r}^{2}\rangle,
\label{rms}
\ee
with
\be
\delta\langle \textbf{r}^{2}\rangle={{1}\over{A+2}}\left({{2A}\over{A+2}}
\langle  \boldsymbol{\lambda}^{2}\rangle+{{1}\over{2}}\langle \boldsymbol{\rho}^{2}\rangle \right).
\label{delta}
\ee
In Eq.(\ref{delta}), $\boldsymbol{\lambda}$ is the distance between the center-of-mass of the two extra-nucleons and the center-of-mass of the core and $\boldsymbol{\rho}$ 
the distance between the two nucleons, 
\bea
\boldsymbol{\lambda}&=&{\dfrac{1}{2}}(\textbf{r}_{1}+\textbf{r}_{2}), \label{lambda}\\
\boldsymbol{\rho}&=&\textbf{r}_{1}-\textbf{r}_2, \label{rho}
\eea
where $\textbf{r}_{1}$ and $\textbf{r}_{2}$ are the coordinates of the two extra-nucleons relative to the center-of-mass of the core. 

The model assumes an inert core plus two correlated neutrons. Therefore the $B(E1)$ for a transition from the 0$^{+}_{1}$ ground state to the 1$^{-}_{f}$ excited state 
is given by the so-called soft dipole strength \cite{suzuki_90}, 
\be
B_{f}(E1)=\left(\dfrac{e_{n}}{e}\right)^{2}\ \left|\langle A+2, 1^{-}_{f}||\sum_{i=1}^{2} r_{i} Y_{1}(\omega_{i})||A+2, 0 \rangle \right|^2,
\ee
where the sum runs over the two extra-neutrons. The contribution to the $E1$ strength comes from only the two extra neutrons with an effective charge,
\be
 e_{n}=\dfrac{-Z e}{A+2}, \label{eff}
\ee
where $Z$ is the number of protons in the core. The expression of $B(E1)$ and more general transition probabilities in the pp-RPA formalism are given in Appendix \ref{trans}. In addition, a simple expression for the sum of $B(E1)$ over all the dipole states can be derived \cite{esbensen_92},
 \be
 \sum_f B_{f}(E1)= \dfrac{3}{\pi}\left(\dfrac{Z}{A+2}\right)^{2}\left\langle \boldsymbol{\lambda}^{2} \right\rangle,
 \label{sumrule}
 \ee
This formula is very useful because it provides a constraint between $\sum_{f}B_{f}(E1)$ and $\boldsymbol{\lambda}$: the $E1$ strength extracted from experiment should not exceed the value obtained in the right hand side of Eq.(\ref{sumrule}), calculated with the experimental value of $\left\langle \boldsymbol{\lambda}^{2} \right\rangle$.

\section{Description of even-even $^{8-12}$Be from a $^{10}$Be core}
\label{result10Be}

In this part, we study even-even $^{8-12}$Be isotopes using the pp-RPA formalism with a $^{10}$Be core. It is well-established that the pure 
Hartree-Fock (HF) approximation fails to reproduce $^{11}$Be properties, in particular the inversion between {1/2}$^+$ and {1/2}$^-$ states \cite{ajzenberg_90}. 
This inversion is due to the coupling with a low energy 2$^{+}$ collective state of the $^{10}$Be core (E$_{x}$=3.36 MeV) characterized by a strong quadrupole 
transition probability ($B(E2;$0$^{+}_{1}$ $\rightarrow$2$^{+}_{1})$=52 e$^{2}$fm$^{4}$) \cite{ajzenberg_84,vinhmau_95,nunes_02}. A phenomenological correction to the one-body potential, simulating the coupling between the neutron and a phonon of the core, has been proposed in Ref.\cite{vinhmau_95}. This new one-body potential is written as
\be
V_{\nu}(r)=V_{WS}(r)+\delta V_{\nu}(r),
\label{ws}
\ee
with
\be
\delta V_{\nu}(r)=16 \ \ a^{2} \alpha_{\nu} \left(\dfrac{df(r)}{dr}\right)^2.
\label{pv}
\ee
In Eq.(\ref{ws}), V$_{WS}$ is a WS potential including central and spin-orbit parts plus a symmetry 
energy term accounting for neutron excess. In Eq.(\ref{pv}), $f(r)$ is the Fermi form factor of the WS potential characterized by 
the diffuseness $a$. The values of parameters of Ref.\cite{pacheco_02} have been adopted. 
Eigen wave functions of this adopted WS potential contain the effect of the coupling between a neutron and a phonon of the core. 
They are no longer pure one-particle wave functions but a mixing of one single neutron 
state coupled to the ground state of the core and of a single neutron state coupled to a phonon of the core. 
In practice, parameters $\alpha_{\nu}$ of Eq.(\ref{pv}) have been fitted from the experimental spectrum of $^{11}$Be \cite{ajzenberg_90} 
for 2s$_{1/2}$, 1p$_{1/2}$ and 1d$_{5/2}$ neutron states and taken equal to zero for the other states. From a technical viewpoint, a 20 fm range box has been used 
to determine eigen wave functions. 

In Table \ref{table1}, results obtained for ground state properties of $^{10,12}$Be are presented. A good agreement for two neutron separation energies S$_{2n}$ in $^{10}$Be and $^{12}$Be is found. The radius of $^{12}$Be calculated using Eq.(\ref{rms}) is also well reproduced. Predictions for the mean distances $\langle\boldsymbol{\lambda}^{2}\rangle^{1/2}$ and $\langle \boldsymbol{\rho}^{2}\rangle^{1/2}$ associated with $\boldsymbol{\lambda}$ and 
$\boldsymbol{\rho}$, Eqs.(\ref{lambda}) and (\ref{rho}), are indicated although no experimental values are yet available. 
\begin{table}[h!]
\caption {Theoretical and experimental values of $S_{2n}$ (MeV) in $^{10}$Be and $^{12}$Be (from Ref.\cite{audi_03}), $\langle \textbf{r}^{2}\rangle^{1/2}$ 
(fm), $\langle \boldsymbol{\lambda}^{2}\rangle^{1/2}$  (fm) and  $\langle\boldsymbol{\rho}^{2}\rangle_{A+2}^{1/2}$ (fm) in $^{12}$Be (from Ref.\cite{ozawa_01}).}
\begin{center}
\begin{tabular}{cccccc}
\hline
\hline
       & $S_{2n}$($^{10}$Be) & $S_{2n}$($^{12}$Be)  & $\langle \textbf{r}^{2}\rangle_{A+2}^{1/2}$ &  $\langle\boldsymbol{\rho}^{2}\rangle^{1/2}$ &$\langle \boldsymbol{\lambda}^{2}\rangle^{1/2}$ \\
\hline
Theory &     8.49    &      3.62       &    2.76                             &             4.89                   &      4.10  \\
\hline
Exp.  &     8.48    &   3.67$\pm$0.01 &    2.59$\pm$0.06                    &               -                      &       -     \\
\hline
\hline
\end{tabular}
\end{center}
\label{table1}
\end{table}

In Table \ref{table3}, the main pp-RPA amplitudes contributing to the ground state wave function of $^{12}$Be are shown. 
From the values of X$_{ab}$ that describe the contribution of a configuration where two neutrons are created in two unoccupied particle states, one sees that the wave function of $^{12}$Be is a mixing of different configurations with the two extra-neutrons mainly in (1p$_{1/2}$)$^2$, (2s$_{1/2}$)$^2$ and (1d$_{5/2}$)$^2$ configurations. The value of $X_{\alpha \beta}$ that stands for two-neutron configurations in hole states is quite large indicating strong correlations in the $^{10}$Be-core with the presence of two particles - two holes configurations. Such a strong mixing is an indication of a breakdown of the N=8 shell closure in $^{12}$Be. This result is supported by experimental data \cite{navin_00} and is consistent with results provided by other models \cite{nunes_02}. 

\begin{table}[h!]
\caption {Main pp-RPA amplitudes of 0$^{+}_{1}$ ground state in $^{12}$Be.}
\begin{center}
\begin{tabular}{cccc|c}
\hline
\hline
                 &      X$_{ab}$            &                  &                  &   X$_{\alpha \beta}$ \\
(1p$_{1/2}$)$^2$ & (1p$_{1/2}$, 2p$_{1/2}$) & (2s$_{1/2}$)$^2$ & (1d$_{5/2}$)$^2$ & (1p$_{3/2}$)$^2$       \\
\hline
   0.76          &        0.31              &      0.50        &   0.43         &    0.57              \\
\hline
\hline
\end{tabular}
\end{center}
\label{table3}
\end{table}
As discussed, results obtained for ground state properties of $^{10}$Be and $^{12}$Be are in an overall good agreement with known experimental data. 
        \begin{figure}[h]
       	\begin{center}
         \vspace{0.9cm}
         \includegraphics[width=0.45\textwidth]{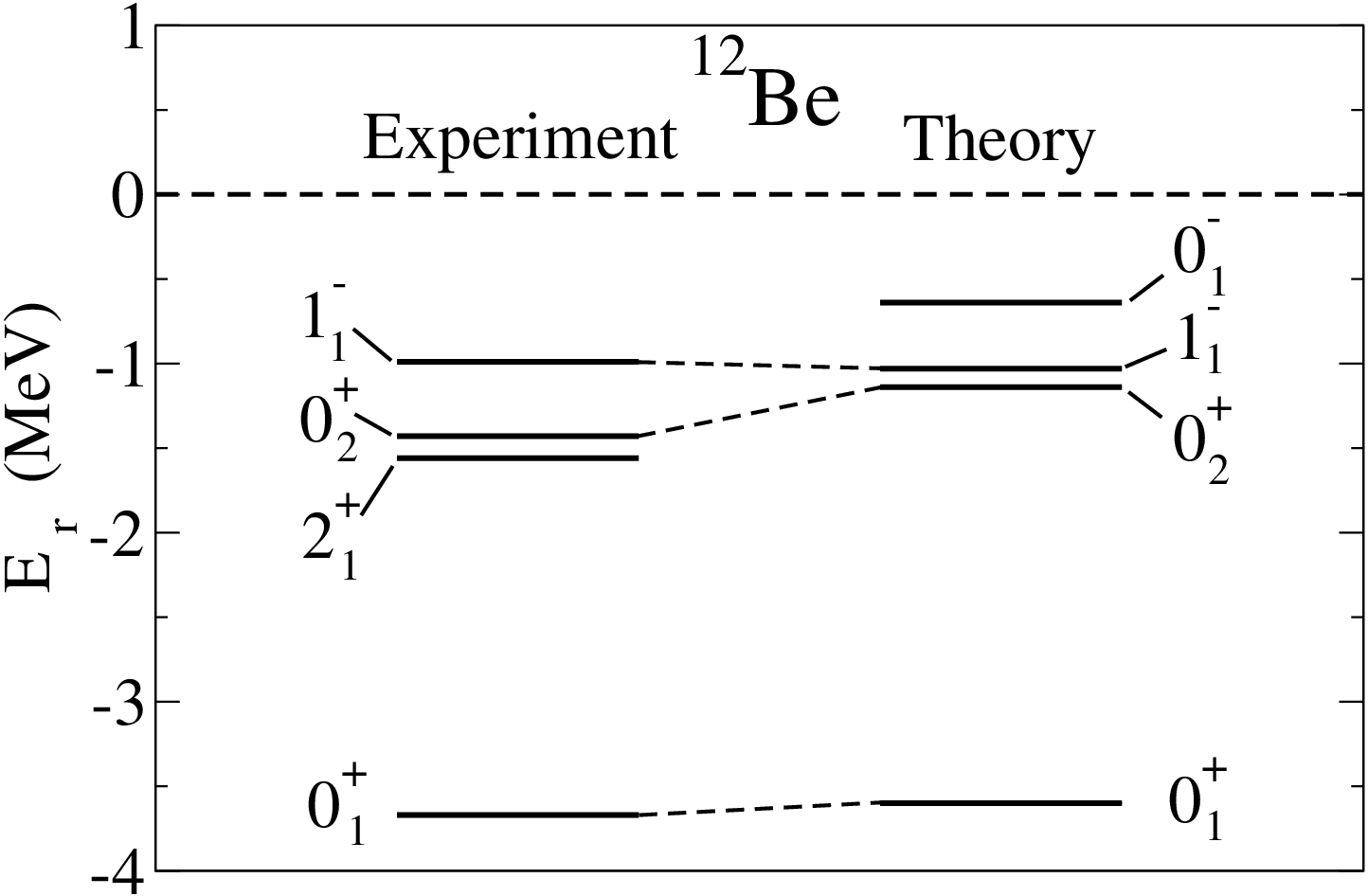}
         \caption{\label{12be_etats} Low lying spectra of $^{12}$Be obtained with pp-RPA compared to experiment.}
       	\end{center}
        \end{figure}
Now we extend our calculation to the description of excited states. We focus on the 0$^{+}$, 0$^{-}$, 1$^-$ and 2$^+$ excited states of 
$^{8}$Be and $^{12}$Be. Calculated excitation energies E$_{x}$ as well as dipole transition probabilities $B(E1)$ are compared
to experimental ones. In practice, the pp-RPA formalism provides the energy E$_{r}$ of a state relative to the two neutrons + core threshold and the two neutron separation energy S$_{2n}$. The excitation energy E$_{x}$ is thus expressed as 
\begin{equation}
E_{x} = E_{r} + S_{2n}.
\label{truc1}
\end{equation}

Concerning $^{8}$Be, our model predicts the absence of low lying 1$^{-}$ and 0$^{+}$ excited states that is in agreement with experimental data. Only a 2$^{+}$ state with an excitation energy of 3.82 MeV is found. This state is experimentally observed at a slightly lower energy of 3.03 MeV. 
\begin{table}[h!]
\caption {Main pp-RPA amplitudes of 0$^{+}_{2}$ state in $^{12}$Be. X$_{\alpha \beta}$ amplitudes are found negligible.}
\begin{center}
\begin{tabular}{cccc}
\hline
\hline
                 &      X$_{ab}$    &                         &                            \\
(1p$_{1/2}$)$^2$ & (2s$_{1/2}$)$^2$ & (1p$_{1/2}$,2p$_{1/2}$) &    (2s$_{1/2}$,3s$_{1/2}$) \\
\hline
   -0.48         &      -0.86       &           0.11          &          0.10              \\
\hline
\hline
\end{tabular}
\end{center}
\label{table3bis}
\end{table}

Results for $^{12}$Be are summarized in Fig.\ref{12be_etats}. The experimental 0$^{+}_{2}$ excited state at E$_{x}$=2.24 MeV \cite{shimoura_03} is well reproduced with 
a theoretical excitation energy of 2.48 MeV. In Table \ref{table3bis}, the main amplitudes corresponding to this state which is mainly built from (2s$_{1/2}$)$^2$ and (1p$_{1/2}$)$^2$ configurations are shown. It can be noted that for 0$^{+}_{2}$ state no effect of correlations in the core, which are represented by 
negligible contributions from X$_{\alpha \beta}$ amplitudes, are found. Looking at both Tables \ref{table3} and \ref{table3bis}, one sees that the model gives a higher contribution of the (2s$_{1/2}$)$^2$ configuration in the 0$^{+}_{2}$ excited state than in the ground state 0$^{+}_{1}$. This trend may be related to the results of Kanungo \textit{et al.} \cite{kanungo_10} where the s-wave spectroscopic factor was equal to 0.28 for the ground state and 0.73 for the 0$^{+}_{2}$ excited state.

We also obtain a 1$^{-}_{1}$ state at E$_{x}$=2.59 MeV with a transition probability of $B(E1$;0$^{+}_{1}$ $\rightarrow$ 1$^{-}_{1})$=0.45 e$^2$ fm$^2$. The 
associated experimental values are E$_{x}$=2.68(3) MeV and $B(E1)$=0.051(13) e$^2$ fm$^2$ \cite{iwasaki_00}. The energy of 1$^{-}_{1}$ state is in agreement with experiment whereas the value of the calculated $B(E1)$ is overestimated by a factor of 10. Sagawa \textit{et al.} \cite{sagawa_01} find a low energy transition strength 
of $B(E1)$=0.063 $e^{2}$ fm$^{2}$ in the context of large shell model calculations using extended wave functions for loosely-bound 1p$_{1/2}$ ans 2s$_{1/2}$ states. In 
our model, as equations are solved in a box, we are free from this kind of correction as our eigen wave functions have already good asymptotic properties. One difference may come from the fact that Sagawa \textit{et al.} renormalize the depth of their HF potential to reproduce half of the empirical two-neutron separation energy. 
Their deduced single particle wave functions are thus much more bound than ours, with certainly a smaller spatial extension. This may explain partly why they obtain a lower $E1$ 
strength. 

\begin{figure}[h]
       	\begin{center}
         \vspace{0.9cm}
         \includegraphics[width=0.45\textwidth]{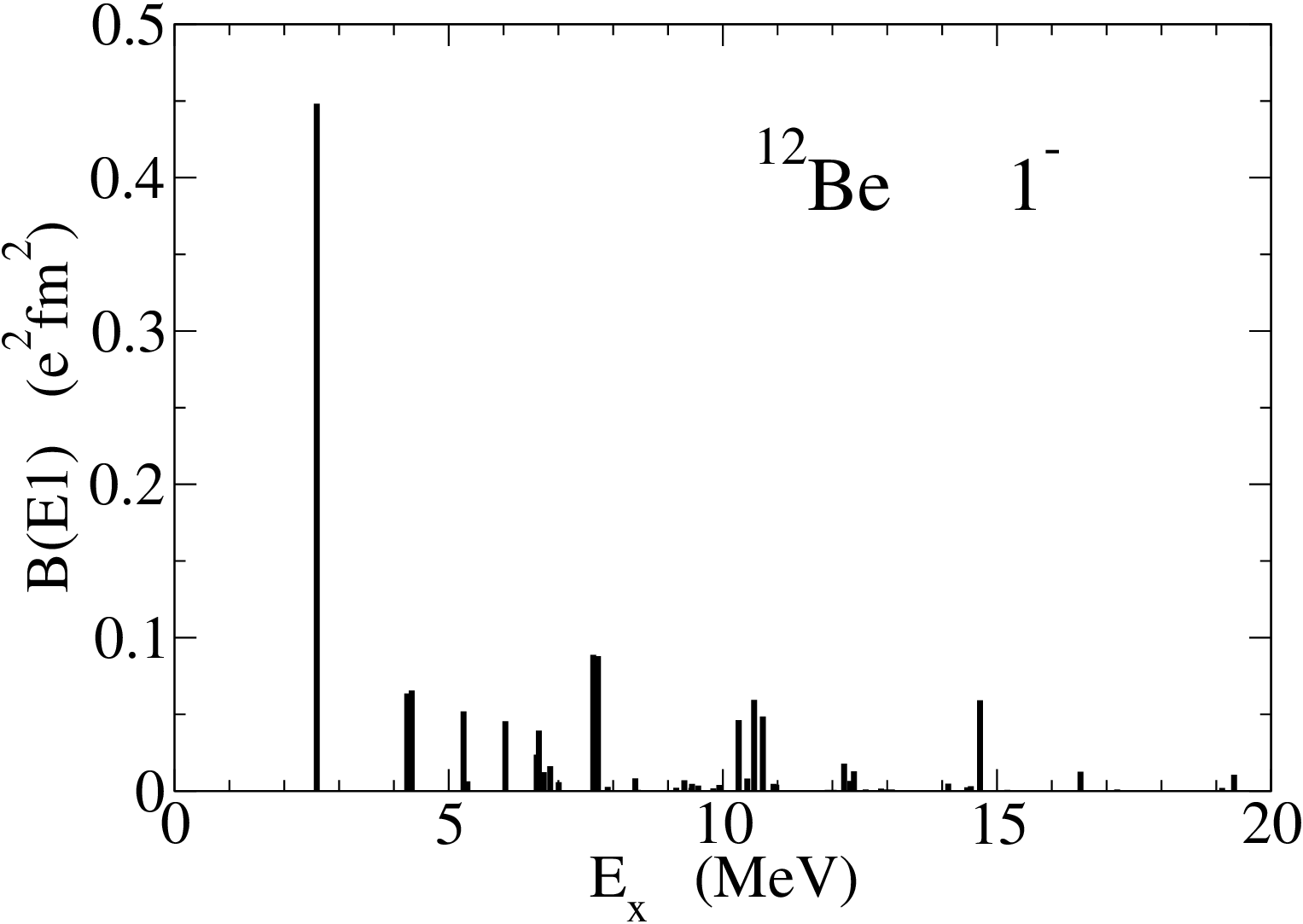}
         \caption{\label{12be_etats_be1}Calculated results of $E1$ strength distribution in $^{12}$Be.}
       	\end{center}
        \end{figure}

In Fig.\ref{12be_etats_be1}, the calculated $E1$ strength distribution in $^{12}$Be is shown. The $E1$ strength associated with the 1$^{-}_{1}$ state gives the largest contribution in our calculation. In Ref.\cite{sagawa_01} where no core is assumed, the contribution of the giant dipole resonance is found in the energy range  E$_{x}$=10-13 MeV. In our calculation only the soft dipole part of the $E1$ strength is accessible. With the help of Eq.(\ref{sumrule}) and the calculated value of $\lambda$, the total deduced $E1$ strength is equal to 1.8 e$^{2}$ fm$^{2}$. Transition probabilities $B(E1)$ as well as radii are two types of observables very sensitive to the content of the wave function. For $^{12}$Be, the ground state is a large mixture of (2s$_{1/2}$)$^{2}$, (1p$_{1/2}$)$^{2}$ and (1d$_{5/2}$)$^{2}$ configurations, as shown in Table \ref{table3}, while the 1$^{-}$ states are nearly pure two-neutron configurations. In particular, the 1$^{-}_{1}$ state is nearly a pure 
(1p$_{1/2}$, 2s$_{1/2}$) configuration. As discussed earlier, the calculated ground state wave function of $^{12}$Be provides a good value for the radius which depends strongly on the values of $\boldsymbol{\lambda}$ and $\boldsymbol{\rho}$. This agreement strongly lends credence to the calculated value of $\sum_{f} B_{f}(E1)$. One may also recall that in $^{11}$Li the calculated 1$^{-}_{1}$ low lying state \cite{bonac_97} was in agreement with later measurement \cite{nakamura_06}, both for excitation energy and transition probability. In addition, our model predicts two higher 1$^{-}$ states characterized by  E$_{x}$=4.24 MeV with $B(E1)$=0.064 $e^{2}$ fm$^{2}$ and E$_{x}$=4.32 MeV with $B(E1)$=0.066 e$^{2}$ fm$^{2}$. These states have not been yet observed experimentally.

Regarding 2$^{+}$ states, our model is unable to reproduce the low lying 2$^{+}_{1}$ state, experimentally found at 2.11 MeV \cite{bernas_62}. However, it predicts 
two additional 2$^{+}$ states at higher excitation energies, 3.86 MeV and 4.59 MeV. These results are coherent with those obtained by Romero-Redondo {\it et al.} \cite{romero_08,romero_08b} where they studied $^{12}$Be within a three-body model with an inert core. They were unable to reproduce the low 2$^{+}_{1}$ state and also predicted a higher state close to our state at 3.8 MeV. Indeed in our model the phenomenological particle-vibration coupling in Eq.(\ref{pv}) takes into account the effect of the 2$^{+}_{1}$ state of $^{10}$Be on the single particle scheme of $^{11}$Be. But the 2$^{+}_{1}$ state of $^{10}$Be itself is not present. As shown by Nunes \textit{et al.} \cite{nunes_02}, in order to explain this state, one has to take into account explicitly the excitations of the $^{10}$Be core.

Concerning the 0$^{-}_{1}$ state, it is obtained in our study with a relative energy of -0.71 MeV (E$_{x}$=2.91 MeV) close to the results of 
Romero-Redondo \textit{et al.} \cite{romero_08} (E$_{x}$=2.5 MeV). It is built from a nearly pure (1p$_{1/2}$, 2s$_{1/2}$) configuration. This 0$^{-}_{1}$ state has not been yet observed even in a recent experiment \cite{kanungo_10}. 

As discussed previously, results obtained for ground states and excited states are found to be in a quite good agreement with experiment, except for the dipole transition probabilities and the 2$^{+}_{1}$ state energy. The pp-RPA approach is able to describe quite well the known spectrum of $^{8}$Be and $^{12}$Be and suggests the presence of higher states which have not been yet observed experimentally.

\section{Description of even-even $^{10-14}$Be from a $^{12}$Be core}
\label{result12Be}
In this part, we are interested in the description of even-even $^{10-14}$Be isotopes using the pp-RPA approach starting from a $^{12}$Be core. Concerning neutron states in the field of $^{12}$Be, the situation is still not clear. A 5/2$^+$ resonance has been first observed at 2.0 MeV above the neutron-$^{12}$Be threshold \cite{ostrowski_92}. This state is interpreted as one neutron in the 1d$_{5/2}$ shell. Later experiments have confirmed this resonance with a relative energy between 2.0 MeV and 2.4 MeV \cite{korsheninnikov_95,belozyorov_98,thoennessen_00,lecouey_04,simon_04,simon_07,kondo}. A lower state has been observed close to 0.3 MeV \cite{thoennessen_00}. A 1/2$^{+}$ assignment, as suggested in Ref.\cite{thoennessen_00}, with a pure neutron s-state implies that the shell order in $^{13}$Be is given by a simple WS potential with an occupied 1p$_{1/2}$ shell. In a recent experiment by Kondo \textit{et al.} \cite{kondo}, they have identified $^{13}$Be resonances of 1/2$^{-}$ with $E_{r}$=0.45 MeV, 1/2$^{+}$ with $E_{r}$=1.17 MeV, and 5/2$^{+}$ with $E_{r}$=2.34 MeV. This experimental result suggests an inversion between the 2s$_{1/2}$ and 1p$_{1/2}$ shells as the one predicted in Ref.\cite{pacheco_02}. In order to clarify those contradictory results, in the following we test two scenarii in $^{13}$Be, 
concerning shell ordering in order to find a scenario that reproduces at best experimental observables for $^{14}$Be. 

In the first scenario, we assume a normal order of shells with a low lying 2s$_{1/2}$ neutron state. The 1p$_{1/2}$ shell is then the last occupied neutron orbital in $^{12}$Be with an energy $\epsilon$(1p$_{1/2}$)=-3.17 MeV given by the measured neutron separation energy in $^{12}$Be \cite{NS}. In the following this scenario is referred as scenario A.

In the second scenario, the inversion between 2s$_{1/2}$ and 1p$_{1/2}$ shells is assumed. Indeed, in $^{12}$Be a 2$^{+}_{1}$ state with an excitation energy of E$_{x}$(2$^{+}_{1}$)=2.6 MeV and a transition probability B(E2;0$^{+}_{1}$ $\rightarrow$ 2$^{+}_{1}$)$\approx$50 e$^{2}$ fm$^{4}$ \cite{iwasaki_00b} is observed close to the one in $^{10}$Be \cite{ajzenberg_84,vinhmau_95}. A similar  effect on the neutron states in $^{13}$Be as the existing one in $^{11}$Be can be expected. Thus assuming that the shell inversion present in $^{11}$Be holds in $^{13}$Be, the last occupied nucleon orbital in $^{12}$Be is the 2s$_{1/2}$ shell with an energy $\epsilon$(2s$_{1/2}$)=-3.17 MeV. The 1p$_{1/2}$ shell is unbound with an energy not established experimentally. In this second scenario the energy of the 1p$_{1/2}$ 
shell is considered as a parameter. The energy of the 1d$_{5/2}$ state is assumed to be $\epsilon$(1d$_{5/2}$)=2.27 MeV, a bit more than the usual one ($\epsilon$(1d$_{5/2}$)=2.0 MeV) but still in agreement with recent experiments \cite{simon_07,kondo}. This result is not in contradiction with experimental knowledge. Indeed, it has been shown in Refs.\cite{simon_07,blanchon_07} that a 1p$_{1/2}$ state above threshold is needed in order to reproduce the experimental neutron-$^{12}$Be spectrum. In the following, this second scenario is referred as scenario B. 

Now, we compare results obtained with the two scenarii with experimental data on even-even $^{10-14}$Be, both for ground state and excited states. 

\begin{table}[h!]
\caption {Theoretical and experimental values of $S_{2n}$ (MeV) in $^{12}$Be and $^{14}$Be (from Ref.\cite{audi_03}), $\langle \textbf{r}^{2}\rangle_{A+2}^{1/2}$ 
(fm), $\langle\boldsymbol{\lambda}^{2}\rangle^{1/2}$ (fm) and $\langle\boldsymbol{\rho}^{2}\rangle^{1/2}$ (fm) in $^{14}$Be (from Refs. \cite{ozawa_01,marques}, $\boldsymbol{\lambda}$ deduced using Eq.(\ref{rms})) in the two cases of non-inversion (A) and inversion (B) of the 2s$_{1/2}$ and 1p$_{1/2}$ shells.}
\begin{center}
\begin{tabular}{cccccc}
\hline
\hline
      & $S_{2n}$($^{12}$Be) & $S_{2n}$($^{14}$Be) &$\langle \textbf{r}^{2}\rangle_{A+2}^{1/2}$ &    $\langle\boldsymbol{\rho}^{2}\rangle^{1/2}$      &
 $\langle\boldsymbol{\lambda}^{2}\rangle^{1/2}$  \\
\hline
  A   &      2.91           &         0.51        &           3.45              &       8.45                             &              5.45                  \\
  B   &      3.71           &         1.29        &           2.91              &       4.56                             &              4.02                  \\
\hline
Exp.  &   3.673$\pm$0.015   &    1.26 $\pm$ 0.01  &       3.10$\pm$0.15         &       5.4$\pm$1.0                      &              4.2$\pm$1.7           \\
\hline
\hline
\end{tabular}
\end{center}
\label{table2}
\end{table}

In scenario A, we first assign a relative energy of 0.3 MeV to the 2s$_{1/2}$ state, according to the experimental suggestion of a low $1/2^{+}$ neutron state from Ref.\cite{thoennessen_00}. In this case all calculated quantities disagree with experimental values. In particular, $^{14}$Be is found under-bound. We then decrease the energy of the 2s$_{1/2}$ state preserving the agreement with experimental data. The energy of the 1d$_{5/2}$ state is fixed at 2 MeV and the one of the 2s$_{1/2}$ state 
at 0.09 MeV (as low as possible ensuring also an unbound $^{13}$Be). Results are summarized in Table \ref{table2}. One sees that, even in that case, it is impossible to describe correctly both the two-neutron separation energies S$_{2n}$ of $^{12}$Be and $^{14}$Be. The $rms$ radius is overestimated and the $rms$ value of $\boldsymbol{\lambda}$ is in the upper part of the experimental error bars. The $rms$ value of $\boldsymbol{\rho}$ is also overestimated by more than 2 fm. It is interesting to note that if the 2s$_{1/2}$ state is bound, results are improved and agree with the $S_{2n}$($^{14}$Be) value given by Descouvemont \textit{et al.} \cite{descouvemont_95,adahchour_95,baye_97}. If the 2s$_{1/2}$ state is unbound and the energy of the d$_{5/2}$ shell is decreased, results closer to experiment 
are found, as in the work of Thompson and Zhukov \cite{thompson_96}. However, these assumptions are not justified since a bound 2s$_{1/2}$ state and a 
1d$_{5/2}$ state below 2 MeV disagree with all experimental measurements. In addition, scenario A has also been studied within a model introducing a core deformation \cite{tarutina_04}. Only for a very high deformation parameter ($\beta > 0.8$), $^{13}$Be has an unbound $1/2^{+}$ ground state and $^{14}$Be a two-neutron separation energy higher than 1 MeV. 

In scenario B, the energy of the 2s$_{1/2}$ state is given by the one-neutron separation energy in $^{12}$Be. The energy of the 1p$_{1/2}$ state is fitted in order to reproduce at best the results for $^{12}$Be and $^{14}$Be. A good agreement with all quantities, including $S_{2n}$($^{12}$Be) is found for a 1p$_{1/2}$ state energy of 0.48 MeV, as shown in Table \ref{table3}. This result for the 1p$_{1/2}$ state energy is in complete agreement with Refs.\cite{kondo,blanchon_07}. Moreover the 3s$_{1/2}$ state is found at 1.33 MeV, corresponding to the 1/2$^{+}$ state with $E_{r}$=1.17 MeV observed in Ref.\cite{kondo}. Note that we find for the 0$^{+}_{1}$ state equivalent results as in Ref.\cite{pacheco_02} where the spin-orbit part of the D1S effective interaction was neglected. 

\begin{table}[h!]
\caption {Main pp-RPA amplitudes for 0$^{+}_{1}$ ground state in $^{14}$Be without (A) and with (B) inversion of 2s$_{1/2}$-1p$_{1/2}$ shells.}
\begin{center}
\begin{tabular}{cccc|cc}
\hline
\hline
          &  X$_{ab}$         &                  &                        &   X$_{\alpha \beta}$ &                           \\
          & (2s$_{1/2}$)$^2$  & (1d$_{5/2}$)$^2$ &                        &   (1p$_{3/2}$)$^2$   &  (1p$_{1/2}$)$^2$         \\
\hline
   A      &  -0.93            &   -0.49          &                        &     0.32             &     0.36                  \\
\hline
          & (1d$_{5/2}$)$^2$  & (1p$_{1/2}$)$^2$ & (1p$_{1/2}$ 2p$_{1/2}$)& (1p$_{3/2}$)$^2$     &  (2s$_{1/2}$)$^2$         \\
\hline
   B      &   -0.56           &    0.70          &       -0.63            &     0.59             &      -0.45                \\
\hline
\hline
\end{tabular}
\end{center}
\label{table4}
\end{table}

The pp-RPA amplitudes of 0$^{+}_{1}$ state in $^{14}$Be obtained for scenarii A and B are presented in Table \ref{table4}. Results depend strongly on the shell inversion hypothesis. As seen from Table \ref{table4}, in scenario A, $X_{\alpha \beta}$ amplitudes are small. This indicates that the core of $^{12}$Be is little affected by two-body correlations, contrary to results reported in Table \ref{table3}. Now, looking at the amplitudes obtained for scenario B, a qualitative agreement with the components of $^{12}$Be (see Table \ref{table3}) is displayed. Indeed, $X_{\alpha \beta}$ amplitude for the configuration (2s$_{1/2}$)$^2$ is quite large indicating that in $^{12}$Be a configuration with two holes in the 2s$_{1/2}$-shell plays an important role. This result is in agreement with the amplitude for the (2s$_{1/2}$)$^{2}$ configuration presented in Table \ref{table3}. Thus concerning amplitudes for $^{14}$Be ground state the scenario B gives solutions more consistent than scenario A. 
        \begin{figure}[h]
        \begin{center}
          \vspace{0.9cm}
         \includegraphics[width=0.45\textwidth]{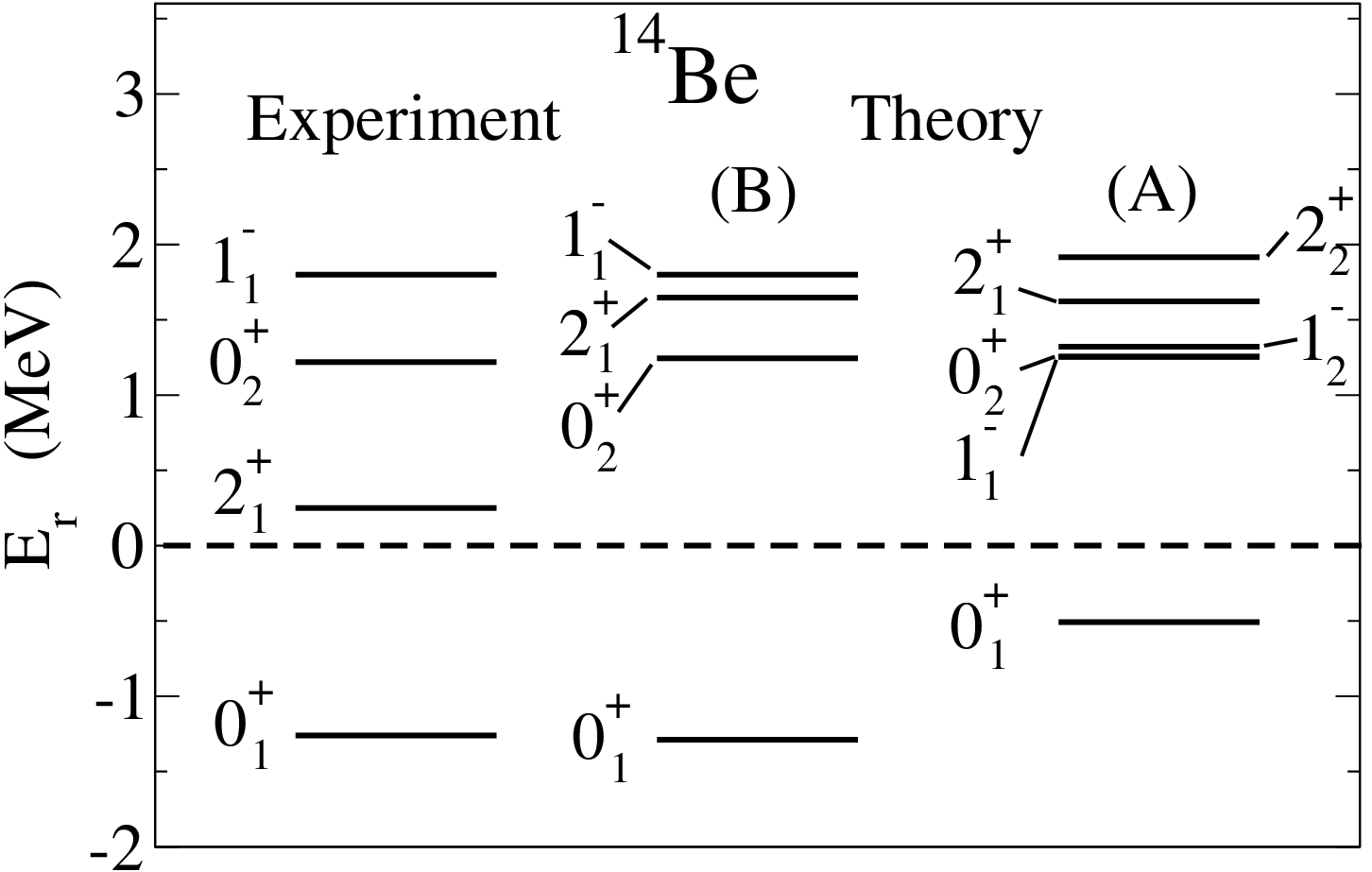}
         \caption{\label{14be_etats} Low lying spectra of $^{14}$Be obtained with pp-RPA without (A) and with (B) inversion in $^{13}$Be compared to experiment.}
        \end{center}
        \end{figure}

In order to understand in more details consequences of scenarii A and B, we study excited states of $^{10}$Be and $^{14}$Be. 

In $^{10}$Be, two 0$^{+}$ and 1$^{-}$ states are observed experimentally at E$_{x}$(0$^{+}_{2}$)=6.18 MeV \cite{ajzenberg_88} and E$_{x}$(1$^{-}_{1}$)=5.96 MeV \cite{ajzenberg_88}, respectively. Two 2$^{+}$ states at E$_{x}$(2$^{+}_1$)=3.37 MeV and E$_{x}$(2$^{+}_2$)=5.96 MeV \cite{ajzenberg_88} are also known from experiment. On the one hand, concerning scenario A, no 1$^{-}$ state is predicted, whereas a 0$^{+}_{2}$ state at 6.31 MeV and a 2$^{+}_{1}$ state at 3.83 MeV are present in the theoretical excited spectrum. Even though those two states seem to be close to experiment, results may not have to be considered satisfactory since they fail to obtain the 
1$^{-}_{1}$ state. On the other hand, scenario B displays a 1$^{-}_{1}$ state at 3.82 MeV and a 0$^{+}_{2}$ state at 3.96 MeV. The agreement is not quantitatively good. However the two states are close to each other as they are found experimentally. This suggests that these states are formed by two neutron holes coupled to the 0$^{+}_{2}$ excited state of $^{12}$Be which has an excitation energy of 2.4 MeV; that is enough to shift the two states at the right energy. 

Results in $^{14}$Be from the two scenarii and comparison with experiment are summarized in Fig.\ref{14be_etats}. Here it is better suited to discuss results in terms of relative energy E$_{r}$ as in scenario A the experimental two-neutron separation energy $S_{2n}$ is not well reproduced. In both scenarii, a good agreement for the relative energy E$_{r}$ of the 0$^{+}_{2}$ excited state is obtained, E$_{r}$=1.26 MeV and 1.24 MeV for scenario A and B respectively. The experimental value is equal to 1.22(18) MeV  \cite{simon_07}. The energies of the 0$^{+}_{2}$ state are close from each other in both scenarii. In Table \ref{table4bis}, the amplitudes for the 
0$^{+}_{2}$ state are displayed for both scenarii. In scenario A, 0$^{+}_{2}$ state is mainly built on (1d$_{5/2}$)$^{2}$, (2s$_{1/2}$, 3s$_{1/2}$) and (2s$_{1/2}$)$^{2}$ configurations. The values of $X_{\alpha\beta}$ indicates the presence of correlations in the core nucleus. In scenario B, the 0$^{+}_{2}$ state is explained with 
mainly (1p$_{1/2}$)$^{2}$ and (1p$_{1/2}$, 2p$_{1/2}$) configurations and the $X_{\alpha\beta}$ are very small. Experimentally only the energy of this state is known. In a future experiment it would be interesting to investigate the spectroscopic factors of this state in order to discriminate between the two scenarii. 
\begin{table}[h!]
\caption {Main pp-RPA amplitudes for the 0$^{+}_{2}$ excited state in $^{14}$Be without (A) and with (B) inversion of 2s$_{1/2}$-1p$_{1/2}$ shells.}
\begin{center}
\begin{tabular}{cccc|cc}
\hline
\hline
          &  X$_{ab}$         &                          &                        &   X$_{\alpha \beta}$ &                    \\
          & (2s$_{1/2}$)$^2$  & (2s$_{1/2}$, 3s$_{1/2}$) &  (1d$_{5/2}$)$^2$      &   (1p$_{3/2}$)$^2$   &  (1p$_{1/2}$)$^2$  \\
\hline
   A      &      0.42         &           0.58           &       -0.73            &        0.27          &     0.27           \\
\hline
          & (1p$_{1/2}$)$^2$  & (1p$_{1/2}$, 2p$_{1/2}$) &                        &   (1p$_{3/2}$)$^2$   &  (2s$_{1/2}$)$^2$  \\
\hline
   B      &       0.73        &       0.64               &                        &        -0.08         &      0.04          \\
\hline
\hline
\end{tabular}
\end{center}
\label{table4bis}
\end{table}

Concerning 1$^{-}$ states in $^{14}$Be, two low lying states with a relative energy $E_{r}$=1.25 MeV and 1.32 MeV are found in scenario A. These values are a bit lower than the experimental one E$_{r}$=1.8$\pm$0.1 MeV obtained by Labiche \textit{et al.} \cite{labiche_01}. $B(E1)$ transition probability between ground state and 
1$^{-}_{1}$ state has not been calculated as both states are not well reproduced in scenario A. 

\begin{figure}[h]
        \begin{center}
          \vspace{0.9cm}
         \includegraphics[width=0.45\textwidth]{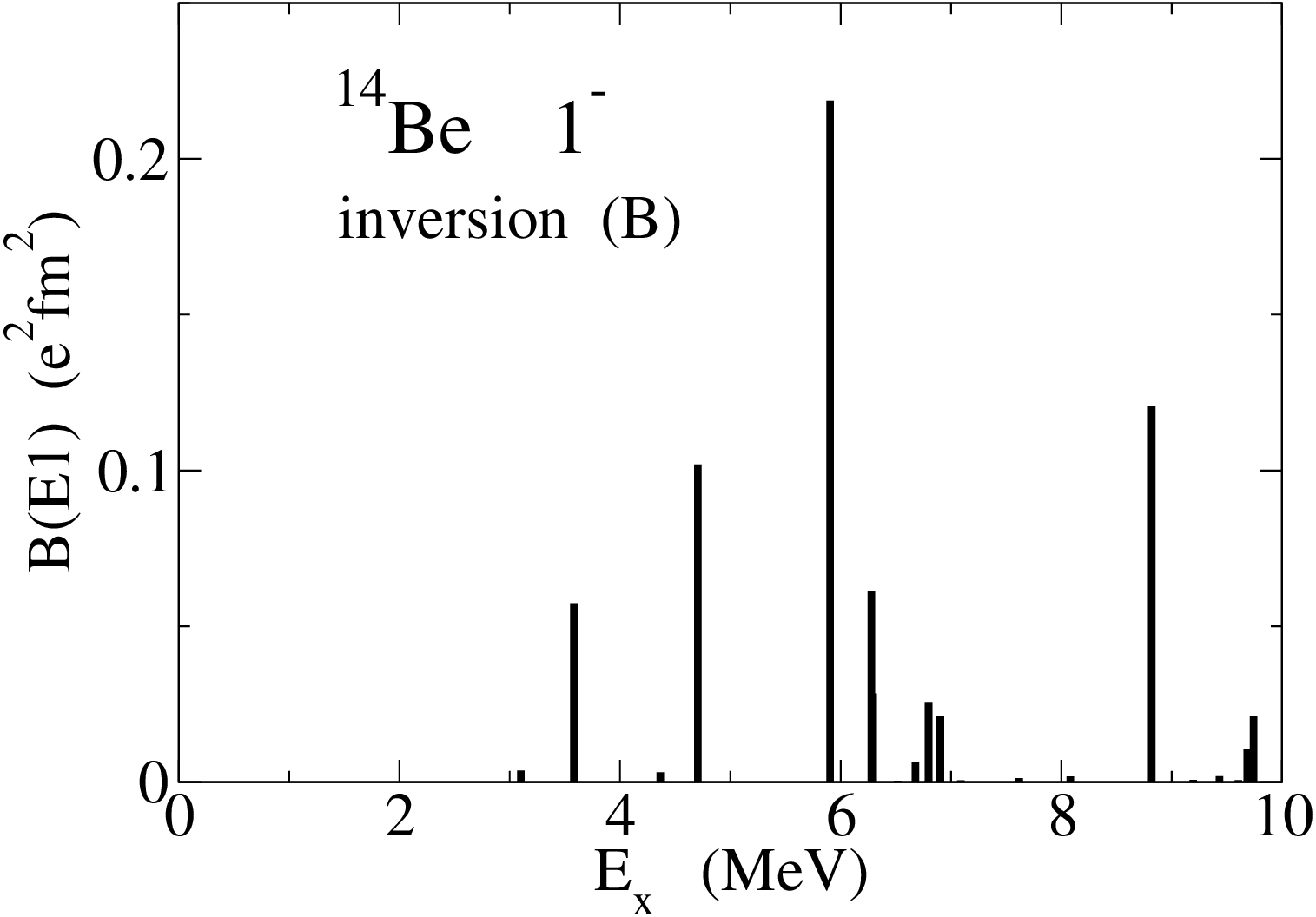}
         \caption{\label{14be_be1}Calculated $E1$ strength distribution in $^{14}$Be for scenario B (with inversion).}
        \end{center}
        \end{figure}

Results appear to be much better in scenario B. In that case, the 1$^{-}_{1}$ state is found at $E_{r}$=1.8 MeV ($E_{x}$=3.1 MeV). Thus ground state properties as well as the energy of the 1$^{-}_{1}$ state are very satisfactory. The $E1$ strength given by our calculation is $B(E1)$=3.7$\times$10$^{-2}$ e$^{2}$fm$^{2}$. It is smaller than the strength obtained by Descouvemont \textit{et al.} of 1.40$\pm$0.40 e$^{2}$fm$^{2}$ in their microscopic cluster model \cite{descouvemont_95,labiche_01}. This difference comes from the fact that they do not assume any inversion in their model. They obtain a s$_{1/2}$ state near threshold in their $^{13}$Be spectrum that enhances the $E1$ strength because of its spatial extension. In Fig.\ref{14be_be1}, the calculated $E1$ strength is shown for the different $1^{-}$ excited states obtained in scenario B. 
The main $E1$ strength is found in the energy-range E$_{x}$=4-7 MeV. The main strength is located at higher excitation energy than seen in $^{12}$Be. The same trend is observed by Sagawa \textit{et al.} \cite{sagawa_01}. In Ref.\cite{forssen_02} Forss\'en \textit{et al.} have extracted the $B(E1)$ distribution from the experimental data of Ref.\cite{labiche_01}. They have found a narrow $B(E1)$ distribution peaked at about E$_{r}$=2 MeV ($E_{x}$=3.3 MeV). The shape of the strength is thus very different from the usual accumulation of $E1$ strength observed at low energy in other Borromean nuclei such as $^{11}$Li and $^{6}$He. Even using a phenomenological model of Coulomb dissociation with a lot of degrees of freedom, they did not manage to fit the shape of the $E1$ strength in $^{14}$Be. This result is a strong indication of a different structure in $^{14}$Be in comparison with other Borromean nuclei. The absence of a low lying neutron s-state in the spectrum of $^{13}$Be and the appearance of a p-state is a possible explanation of the unusual shape of the soft $E1$ strength. In that sense results obtained for $B(E1)$ are in agreement with scenario B. 

Concerning the $E1$ transition probability strength, a value of 0.17 e$^{2}$fm$^{2}$ is found integrated below $E_{r}$=3.2 MeV ($E_{x}$=4.5 MeV). This result seems low compared with the strength of 1.40$\pm$0.40 e$^{2}$fm$^{2}$ extracted by Forss\'en \textit{et al.} Using the sum rule formula of Eq.(\ref{sumrule}) with the calculated $rms$ value of $\boldsymbol{\lambda}$, we obtain a value of 1.26 e$^{2}$fm$^{2}$ in agreement with the one of Ref.\cite{forssen_02}. The sum rule gives the $E1$ transition probability  strength integrated over the whole spectrum. We find that the $E1$ strength extracted below E$_{r}$=3.2 MeV ($E_{x}$=4.5 MeV) is larger than the sum rule as found in Ref.\cite{forssen_02}. Then the results deduced from the experiment of Labiche \textit{et al.} may seem doubtful. 

Concerning the 2$^{+}_{1}$ state in $^{14}$Be, it was first observed by Bohlen \textit{et al.} \cite{bohlen_95} with an excitation energy of E$_{x}$=1.59(11) MeV (E$_{r}$=0.25(6) MeV). Then, this state was confirmed by Korsheninnikov \textit{et al.} \cite{korsheninnikov_95}. In a more recent experiment, Sugimoto \textit{et al.} \cite{sugimoto_07} found the 2$^{+}_{1}$ state at E$_{x}$=1.54(13) MeV (E$_{r}$=0.28$\pm$0.01 MeV). As already discussed for $^{12}$Be, this state is absent from our model as it is interpreted as an excitation of the $^{12}$Be core. However, as shown in Fig.\ref{14be_etats}, a 2$^{+}_{1}$ state at E$_{r}$=1.6 MeV is obtained in both scenarii. 

We think that further investigations are needed both on the theoretical and experimental parts in order to 
achieve a fully consistent description of $^{13}$Be and $^{14}$Be. In our study scenario B reproduces ground state 
as well as excited state properties, except the 2$^{+}_{1}$ state. These results are a strong indication on the validity of scenario B which assumes an inversion between 2s$_{1/2}$ and 1p$_{1/2}$ shells in $^{13}$Be.

\section{Conclusions}
\label{concl}

We have employed the pp-RPA approach to describe even-even $^{8-14}$Be isotopes from either a $^{10}$Be or a $^{12}$Be core. A WS potential corrected by a 
phenomenological particle-vibration coupling for the neutron-core interaction and the D1S Gogny force for the neutron-neutron interaction have been employed.

Starting from the experimental spectrum of $^{11}$Be and a $^{10}$Be core, our approach has been able to provide ground state properties as well as excitation energies of the 0$^{+}_{2}$ and 1$^{-}_{1}$ states in $^{12}$Be. As pp-RPA model assumes an inert core, it fails in describing the 2$^{+}_{1}$ state that is probably built on 
an excited $^{10}$Be core. This issue could be cured in a further work introducing explicitly excitations of the core. The calculated $B(E1)$ transition probability of 
the 1$^{-}_{1}$ state overestimates the experimental value. 

Concerning the description of the most exotic Beryllium isotopes, we have applied the same method with a $^{12}$Be core. In that case, $^{13}$Be spectrum has been needed as input of the calculation. Then, two scenarii have been tested: \textit{i)} scenario A with a normal shell order in $^{13}$Be, \textit{ii)} scenario B with a shell inversion similar to $^{11}$Be. Scenario A leads to several inconsistencies and is unable to reproduce neither ground state nor  excited state properties of $^{14}$Be. In scenario B, the low energy of the 1p$_{1/2}$ neutron state in the field of $^{12}$Be have been used as free parameter. To fix this parameter two constraints have been required. Firstly, the two-neutron separation energy in $^{12}$Be should be in agreement with the experimental value. Secondly, $^{10}$Be and $^{14}$Be ground and excited states should be in agreement with available data. These two different constraints have led to the conclusion of a 2s$_{1/2}$-1p$_{1/2}$ shell inversion in $^{13}$Be as observed in $^{11}$Be and $^{10}$Li. Such an assumption has implied the existence of a $1/2^{-}$ state in $^{13}$Be with an energy around 0.48 MeV, close to threshold. This state seems to have been observed in a recent experiment at RIKEN \cite{kondo}.

The study of the two scenarii argues in favor of the conclusion of Refs.\cite{labiche_99,pacheco_02}: the 2s$_{1/2}$-1p$_{1/2}$ shell inversion is present in $^{13}$Be. The present situation looks like the one of $^{10}$Li when theoretical studies on $^{11}$Li had predicted the necessity to have an inversion in $^{10}$Li \cite{thompson_94,vinhmau_96}, before it was confirmed experimentally \cite{aoi_97,simon_99}. 

\appendix
\section{pp-RPA model and related algebra}
\label{pprpa}

In order to simplify notations, we do not specify the total angular momentum in the pair creation and annihilation operators (see Eq.(\ref{aa1})), except when necessary. One obtains the following relations for amplitudes $X$ and $Y$,
\begin{align}
\sum_{ab}& X_{ab}^{(N)}X_{ab}^{(N')}-\sum_{\alpha \beta}X_{\alpha \beta}^{(N)}X_{\alpha \beta}^{(N')}&=&\delta_{NN'} \\
\sum_{ab}& Y_{ab}^{(M)}Y_{ab}^{(M')}-\sum_{\alpha \beta}Y_{\alpha \beta}^{(M)}Y_{\alpha \beta}^{(M')}&=&-\delta_{MM'} \\
\sum_N& X^{(N)}_{ab}X^{(N)}_{kl} -\sum_M  Y^{(M)}_{ab}Y^{(M)}_{kl}&=&\delta_{ak} \delta_{bl} \\
\sum_N& X^{(N)}_{\alpha \beta}X^{(N)}_{\kappa \lambda} -\sum_M  Y^{(M)}_{\alpha \beta}Y^{(M)}_{\kappa \lambda}&=&-\delta_{\alpha \kappa} \delta_{\beta \lambda}.
\end{align}
We see from Eqs.(\ref{e}) and (\ref{e2}) that the lowest eigenvalues among the $N$ and $M$ solutions  are directly related to the two-neutron separation energy in the $(A+2)$ and $A$ nuclei, $S_{2n}(A+2)$ and $S_{2n}(A)$ respectively. Indeed if N$_0$ and M$_0$ are the lowest energies of the two series of eigenstates, then
\bea
S_{2n}(A)&=&\Omega_{M_0} \\
S_{2n}(A+2)&=& -\Omega_{N_0}.
\eea
To determine the wave functions of the two systems, we introduce two operators  ${\cal{Q}}^\dagger_N$ and ${\cal{Q}}^\dagger_M$ related to pair creation and annihilation (see Eq.(\ref{aa1})). They are built in such a way that the core nucleus plays the role of a vacuum. They verify the following equations,
\bea
{\cal{Q}}_N|A,\tilde 0\rangle &=& 0 \\
{\cal{Q}}_M|A,\tilde 0\rangle &=& 0.
\eea
Moreover, wave functions of $(A+2)$ and $(A-2)$ systems are expressed as 
\bea
|A+2,N\rangle &=& {\cal{Q}}_{N}^{\dagger}|A,\tilde 0\rangle    \label{qq1}\\
|A-2,M \rangle&=&{\cal{Q}}^{\dagger}_M|A,\tilde 0\rangle  \label{qq2}.
\eea
Requiring the orthonormalization of the wave function, and assuming the quasiboson approximation \cite{ring}, 
\bea
\left\langle A,\tilde 0 |\left[ {\cal{A}}_{kl},{\cal{A}}^{\dagger}_{mn}\right]|A,\tilde 0\right\rangle  &\approx& \delta_{km}\delta_{ln} \\
\left\langle A,\tilde 0|\left[ {\cal{A}}_{\kappa \lambda},{\cal{A}}^{\dagger}_{\mu \nu}\right]|A,\tilde 0 \right\rangle  &\approx& -\delta_{\ka\mu}\delta_{\la \nu}, 
\eea
a relation between the operators ${\cal{Q}}$ (see Eqs.(\ref{qq1}) and (\ref{qq2})) and ${\cal{A}}$ (see Eq.(\ref{aa1})) can be established. It reads
\bea
{\cal{Q}}^{\dagger}_N&=&\sum_{k \leq l}X_{kl}^{(N)} {\cal{A}}_{kl}^{\dagger}-\sum_{\kappa \leq \lambda} X_{\kappa \lambda}^{(N)} {\cal{A}}_{\kappa \lambda}^{\dagger} \label{truc2}\\
{\cal{Q}}^{\dagger}_M&=&\sum_{\kappa \leq \lambda} Y_{\kappa \lambda}^{(M)} {\cal{A}}_{\kappa \lambda}-\sum_{k \leq l} Y_{kl}^{(M)} {\cal{A}}_{kl}, \label{truc3}
\eea
with the following properties, 
\bea
\left[{\cal{Q}}_N,{\cal{Q}}^{\dagger}_{N'}\right] &=& \delta_{N N'} \\
\left[{\cal{Q}}_M,{\cal{Q}}^{\dagger}_{M'}\right]  &=& \delta_{M M'}.
\eea
All other commutators between two operators ${\cal{Q}}^{\dagger}$ and ${\cal{Q}}$ are equal to zero. Inverting Eqs.(\ref{truc2}) and (\ref{truc3}), 
\begin{align}
& {\cal{A}}^{\dagger}_{kl}=\sum_{N} X^{(N)}_{kl} {\cal{Q}}^{\dagger}_N +\sum_{M} Y^{(M)}_{kl} {\cal{Q}}_M \\
& {\cal{A}}^{\dagger}_{\ka \la}=\sum_N X^{(N)}_{\ka \la} {\cal{Q}}^{\dagger}_N +\sum _M Y^{(M)}_{\ka \la} {\cal{Q}}_M.
\end{align}

\section{Average value of one and two-body operators}
\label{radius}
 
The calculation of the $rms$ radius of Eq.(\ref{rms}) needs the calculation of the average value of one- and two-body operators 
on the system formed by the two extra-nucleons. Then, given an operator $\widehat{F}$, the following evaluation is required 
\be
\langle \widehat{F} \rangle=\langle A+2,0|\widehat{F}|A+2,0 \rangle - \langle A,\tilde 0|\widehat{F}|A,\tilde 0 \rangle. \label{F}
\ee 
The wave function $|A+2,0\rangle$ corresponds to the ground state,
\be
|A+2, 0\rangle={\cal{Q}}_{0}^{\dagger}|A,\tilde 0\rangle, 
\ee
with
\be
{\cal{Q}}_{0}^{\dagger}=\sum_{a \leq b}X_{ab}^{(0)}{\cal{A}}_{ab}^{\dagger}-\sum_{\alpha \leq\beta}
X_{\alpha\beta}^{(0)}{\cal{A}}_{\alpha\beta}^{\dagger},
\ee
so that Eq.(\ref{F}) can be transformed into
\bea
\langle \widehat{F} \rangle&=&\langle A,\tilde
0|{\cal{Q}}_{0}\widehat{F}{\cal{Q}}_{0}^{\dagger}|A,\tilde 0\rangle-\langle A,\tilde0|\widehat{F}|A,\tilde 0\rangle  \nonumber\\
&=& \langle A,\tilde 0|{\cal{Q}}_{0}[\widehat{F},{\cal{Q}}_{0}^{\dagger}]|A,\tilde 0 \rangle. \label{truc4}
\eea
The formula of Eq.(\ref{truc4}) can be applied to any one-body $\widehat{F_{1}}$ or two-body $\widehat{F_{2}}$ operators. 
For a one-body operator, one gets
\bea
\langle \widehat{F_1} \rangle&=&2\sum_{m < n}\langle m|F_1|n\rangle \sum_{a}X_{an}^{(0)}X_{am}^{(0)}\nonumber \\
&-&2\sum_{\mu < \nu}\langle \mu|F_1|\nu\rangle \sum_{\alpha} X_{\alpha\mu}^{(0)}X_{\alpha\nu}^{(0)},
\label{onebody}
\eea
and for a two-body operator, one gets
\bea
\langle \widehat{F_2} \rangle&=&\sum_{k<l,m<n}\langle kl|F_2|\widetilde{mn}\rangle X_{mn}^{(0)}X_{kl}^{(0)}\nonumber \\
&+&2 \sum_{k<l,\mu<\nu}\langle kl|F_2|\widetilde{\mu\nu} \rangle X_{\mu\nu}^{(0)}X_{kl}^{(0)}\nonumber \\
&+&\sum_{\kappa<\lambda,\mu<\nu} \langle \kappa\lambda|F_2|\widetilde{\mu\nu} \rangle X_{\mu\nu}^{(0)}X_{\kappa\lambda}^{(0)}.
\label{twobody}
\eea

\section{Transition amplitudes in the $A+2$ nucleus}
\label{trans}

In this Appendix, we consider a transition between the ground state ($N$=0) and an excited state ($N$ $\neq 0$ ) of the ($A+2$) 
nucleus through a one-body operator ($\widehat{F_{1}}$), as for example electromagnetic ones. We do not specify angular momentum couplings 
for simplicity. The amplitude for such a transition is given by
\be
{\cal{M}}(0\rightarrow N)=\sum_{i \leq j}\langle i|F_{1}|j \rangle \langle A,\tilde 0|{\cal{Q}}_{N}a_{i}^{\dagger}a_{j}{\cal{Q}}_{0}^{\dagger}|A,\tilde 0\rangle.
\ee
The sum over $i$ and $j$ runs over all nucleon states, occupied or unoccupied. We can then rewrite $\cal{M}$ as
\be
{\cal{M}}(0\rightarrow N )=\sum_{i \leq j}\langle i|F_{1}|j \rangle\langle A,\tilde 
0|{\cal{Q}}_{N}[a_{i}^{\dagger}a_{j},{\cal{Q}}_{0}^{\dagger}]|A,\tilde 0\rangle.
\ee
Then, using the following relation,
\bea
\langle A,\tilde 0|{\cal{Q}}_{N}{\cal{A}}_{ij}^{\dagger}|A,\tilde 0\rangle
&=&\langle A+2, N|{\cal{A}}_{ij}^{\dagger}|A,\tilde 0\rangle \nonumber \\
&=&X_{ij}^{(N)},
\eea
one obtains, 
\begin{multline}
{\cal{M}}(0\rightarrow  N)=\sum_{a_{0} \leq b_{0}, a_{1} \leq b_{1}}X_{a_{1}b_{1}}^{(N)} X_{a_{0}b_{0}}^{(0)}\\
\times [\langle a_{1}|F_{1}|a_{0}\rangle \delta_{b_1b_0}+\langle b_{1}|F_{1}|b_{0}\rangle \delta_{a_0 a_1}]  \\
-\sum_{\alpha_{0} \leq \beta_{0}, \alpha_{1} \leq \beta_1}X_{\alpha_1 \beta_1}^{(N)}X_{\alpha_{0}\beta_{0}}^{(0)} \\
\times [\langle \alpha_{1}|F_{1}|\alpha_{0}\rangle \delta_{\beta_0 \beta_1}+\langle \beta_{1}|F_{1}| \beta_{0}\rangle \delta_{\alpha_0 \alpha_1}].
\end{multline}
In the particular case of an $E1$ transition between $0^{+}_{1}$ ground state and the 1$^{-}_{1}$ excited state and within the hypothesis of an inert core surrounded 
by two extra-neutrons, $F_{1}$ is the electric dipole moment exciting the soft dipole mode \cite{suzuki_90}, 
\be
F_{1}\equiv \frac{e_{n}}{e}\sum_{i=1}^{2} r_{i} Y_{1}^{\mu}(\omega_{i}),
\ee
where $e_{n}$ is the neutron effective charge defined in Eq.(\ref{eff}). Thus the soft $E1$ strength is 
\be
B(E1)=|{\cal{M}}(0^{+}_{1}\rightarrow 1^{-}_{1})|^{2}.
\ee


\begin{thebibliography}{90}
\bibitem{labiche_99} M. Labiche, F. M. Marqu\'es, O. Sorlin, and N. Vinh Mau, Phys. Rev. C {\bf 60}, 027303 (1999).
\bibitem{pacheco_02} J. C. Pacheco and N. Vinh Mau, Phys. Rev. C {\bf 65}, 044004 (2002).
\bibitem{tanihata_85} I. Tanihata, \textit{et al.}, Phys. Rev. Lett. {\bf 55}, 2676 (1985); Phys. Lett. B {\bf 206}, 592 (1988).
\bibitem{chartier_01} M. Chartier, \textit{et al.}, Phys. Lett. B {\bf 510}, 24 (2001).
\bibitem{berger_91} J. F. Berger, M. Girod, and D. Gogny, Comp. Phys. Comm. {\bf 63}, 365 (1991).
\bibitem{decharge_80} J. Decharg\'e and D. Gogny, Phys. Rev. C {\bf 21} 1568 (1980).
\bibitem{shimoura_03} S. Shimoura, {\it et al.}, Phys. Lett. B \textbf{560}, 31 (2003).
\bibitem{lecouey_04} J. L. Lecouey, Few-Body Systems \textbf{34}, 21 (2004). 
\bibitem{simon_04} H. Simon, \textit{et al.}, Nucl. Phys. \textbf{A734}, 323 (2004). 
\bibitem{simon_07} H. Simon, {\it et al.}, Nucl. Phys. {\bf A791}, 267 (2007).
\bibitem{sugimoto_07} T. Sugimoto, {\it et al.}, Phys. Lett. B {\bf 356}, 160 (2007).
\bibitem{kondo} Y. Kondo, {\it et al.}, Phys. Lett. B {\bf 690}, 245 (2010).
\bibitem{kanungo_10} R. Kanungo, {\it et al.}, Phys. Lett. B \textbf{682}, 391 (2010).
\bibitem{vinhmau_95} N. Vinh Mau, Nucl. Phys. {\bf A592}, 33 (1995). 
\bibitem{girod_83} M. Girod and B. Grammaticos, Phys. Rev. C {\bf 27}, 2317 (1983).
\bibitem{ajzenberg_90} F. Ajzenberg-Selove, Nucl. Phys. \textbf{A506}, 1 (1990).
\bibitem{ring} P. Ring and P. Schuck, \textit{The Nuclear Many-Body Problem} (Springer, New York, 1980).
\bibitem{mulhall_67} W. J. Mulhall, R. J. Liotta, J. A. Evans, and R. P. Perazzo, Nucl. Phys. {\bf A93}, 261 (1967).
\bibitem{vinhmau_96} N. Vinh Mau and J. C. Pacheco,  Nucl. Phys. {\bf A607},163 (1996).
\bibitem{suzuki_90} Y. Suzuki and Y. Tosaka, Nucl. Phys. {\bf 517}, 599 (1990). 
\bibitem{nunes_02} F. M. Nunes, I. J. Thompson, and J. A. Tostevin, Nucl. Phys. \textbf{A703}, 593 (2002).
\bibitem{esbensen_92} H. Esbensen and G. F. Bertsch, Nucl. Phys. {\bf A542}, 310 (1992).
\bibitem{ajzenberg_84} F. Ajzenberg-Selove, Nucl. Phys. \textbf{A413}, 1 (1984).
\bibitem{audi_03} G. Audi, A. H. Wapstra, and C. Thibault, Nucl. Phys. \textbf{A729}, 337 (2003).
\bibitem{ozawa_01} A. Ozawa, T. Suzuki, and I. Tanihata, Nucl. Phys. {\bf A693}, 32 (2001).
\bibitem{navin_00} A. Navin, {\it et al.}, Phys. Rev. Lett. {\bf 85}, 266 (2000).
\bibitem{iwasaki_00} H. Iwasaki, {\it et al.}, Phys. Lett. B {\bf 491}, 8 (2000).
\bibitem{sagawa_01} H. Sagawa, T. Suzuki, H. Iwasaki, and M. Ishihara, Phys. Rev. C \textbf{63}, 034310 (2001).
\bibitem{bonac_97} A. Bonaccorso and N. Vinh Mau,  Nucl. Phys. {\bf A615}, 245 (1997).
\bibitem{nakamura_06} T. Nakamura, {\it et al.}, Phys. Rev. Lett. {\bf 96}, 252502 (2006). 
\bibitem{bernas_62} M. Bernas, J. C. Peng, and N. Stein, Phys. Lett. B {\bf 116}, 7 (1962).
\bibitem{romero_08} C. Romero-Redondo, E. Garrido, D. V. Fedorov, and A. S. Jensen, Phys. Lett. B {\bf 660}, 32 (2008).
\bibitem{romero_08b} C.Romero-Redondo, E. Garrido, D. V. Fedorov, and A. S. Jensen, Phys. Rev. C \textbf{77}, 054313 (2008).
\bibitem{ostrowski_92} A. N. Ostrowski, \textit{et al.}, Z. Phys. A {\bf 343}, 489 (1992).
\bibitem{korsheninnikov_95} A. A. Korsheninnikov, \textit{et al.}, Phys. Lett. B {\bf 343}, 53 (1995). 
\bibitem{belozyorov_98} A. V. Belozyorov, \textit{et al.}, Nucl. Phys. {\bf A636}, 419 (1998).
\bibitem{thoennessen_00} M. Thoennessen, S. Yokoyama, and P. G. Hansen, Phys. Rev. C \textbf{63}, 014308 (2000).
\bibitem{NS} http://nucleardata.nuclear.lu.se/database/masses/
\bibitem{iwasaki_00b} H. Iwasaki, {\it et al.}, Phys. Lett. B \textbf{481}, 7 (2000).
\bibitem{blanchon_07} G. Blanchon, A. Bonaccorso, D. M. Brink, and A. Garc\'\i a-Camacho, Nucl. Phys. {\bf A784}, 49 (2007).
\bibitem{marques} F. M. Marqu\'es, {\it et al.}, Report No. LPC Caen 99-13.
\bibitem{descouvemont_95} P. Descouvemont, Phys. Rev. C {\bf 52}, 704 (1995).
\bibitem{adahchour_95} A. Adahchour, D. Baye, and P. Descouvemont, Phys. Lett. B {\bf 356}, 445 (1995).
\bibitem{baye_97} D. Baye, Nucl. Phys. {\bf A627}, 305 (1997).
\bibitem{thompson_96} I. J. Thompson and M. V. Zhukov, Phys. Rev. C {\bf 53}, 708 (1996).
\bibitem{tarutina_04} T. Tarutina, I. J. Thompson, and J. A. Tostevin, Nucl. Phys. {\bf A733}, 53 (2004).
\bibitem{ajzenberg_88} F. Ajzenberg-Selove, Nucl. Phys. \textbf{A490}, 1 (1988).
\bibitem{labiche_01} M. Labiche, \textit{et al.}, Phys. Rev. Lett. {\bf 86}, 600 (2001).
\bibitem{forssen_02} C. Forss\'en, V. D. Efros, and M. V. Zhukov, Nucl. Phys. {\bf A706}, 48 (2002).
\bibitem{bohlen_95} H. G. Bohlen, \textit{et al.}, Nucl. Phys. \textbf{A583}, 775 (1995). W. von Oerzen, \textit{et al.}, Nucl. Phys. \textbf{A588}, 129c (1995).
\bibitem{thompson_94} I. J. Thompson and M. V. Zhukov, Phys. Rev. C \textbf{49}, 1904 (1994).
\bibitem{aoi_97} N. Aoi, \textit{et al.}, Nucl. Phys. \textbf{A616}, 181 (1997).
\bibitem{simon_99} H. Simon, \textit{et al.}, Phys. Rev. Lett. {\bf 83}, 496 (1999).
\end{thebibliography}
\end{document}